\documentclass[useAMS,usenatbib]{mn2e}

\usepackage{graphicx,amssymb,bm}
\usepackage{subfig,xfrac}
\usepackage{float, Abbrev}
\usepackage[fleqn]{amsmath}  
\usepackage{color}
\usepackage{aasmacros}
\usepackage[draft]{hyperref}

\usepackage{multirow}

\newcommand{\be}{\begin{equation}}
\newcommand{\ee}{\end{equation}}
\newcommand{\ba}{\begin{eqnarray}}
\newcommand{\ea}{\end{eqnarray}}

\def\simlt{\lower.5ex\hbox{$\; \buildrel < \over \sim \;$}}

\newcommand{\fig}{\begin{figure} \begin{center}}
\newcommand{\efig}{\end{center}\end{figure} }
\newcommand{\figs}{\begin{figure*}\begin{minipage}{180mm} \begin{center}}
\newcommand{\efigs}{\end{center}\end{minipage}\end{figure*} }
\def\simgt{\lower.5ex\hbox{$\; \buildrel > \over \sim \;$}}

\def\ie{{\it i.e.}}

\usepackage{graphicx} 

\title[WDM Flux ratios]{Exploiting flux ratio anomalies to probe warm dark matter in future large scale surveys}

\author[D. Harvey et al]
{David Harvey$^{1}$\thanks{e-mail: {\tt david.harvey@epfl.ch}}, Wessel Valkenburg$^{2}$, Amelie Tamone$^{3}$, Alexey Boyarsky$^{1,2}$,
 \newauthor Frederic Courbin$^{3}$ and Mark Lovell$^{4,5}$  \\
$^{1}$Instituut-Lorentz for Theoretical Physics, Universiteit Leiden, Niels Bohrweg 2, Leiden, The Netherlands \\
$^{2}$Institute of Physics, Laboratory for Particle Physics and Cosmology (LPPC), \\
\phantom{$^{2}$}\'Ecole Polytechnique F\'ed\'erale de Lausanne, CH-1015 Lausanne, Switzerland \\
$^{3}$Laboratoire d'Astrophysique, EPFL, Observatoire de Sauverny, 1290 Versoix, Switzerland \\
$^{4}$Center for Astrophysics and Cosmology, Science Institute, University of Iceland, Dunhagi 5, 107 Reykjavik, Iceland \\
$^{5}$Institute for Computational Cosmology, Durham University, South Road, Durham DH1 3LE, UK }

\begin{document}

\date{Accepted ---. Received ---; in original form \today.}

\pagerange{\pageref{firstpage}--\pageref{lastpage}} \pubyear{2017}

\maketitle

\label{firstpage}

\begin{abstract}
Flux ratio anomalies in strong gravitationally lensed quasars constitute a unique way to probe the abundance of non-luminous dark matter haloes, and hence the nature of dark matter. In this paper we identify double imaged quasars as a {\it statistically} efficient probe of dark matter, since they are 20 times more abundant than quadruply imaged quasars. Using {\it N}-body simulations that include realistic baryonic feedback, we measure the full distribution of flux ratios in doubly imaged quasars for cold (CDM) and warm dark matter (WDM) cosmologies. Through this method, we fold in two key systematics -- quasar variability and line-of-sight structures. 
We find that WDM cosmologies predict a $\sim6$~per~cent difference in the  cumulative distribution functions of flux ratios relative to CDM, with CDM predicting many more small ratios. 
Finally, we estimate that $\sim 600$ doubly imaged quasars will need to be observed in order to be able to unambiguously discern between CDM and the two WDM models studied here. Such sample sizes will be easily within reach of future large scale surveys such as Euclid. 
In preparation for this survey data we require discerning the scale  of the uncertainties in modelling lens galaxies and their substructure in simulations, plus a strong understanding of the selection function of observed lensed quasars. 
\end{abstract}

\begin{keywords}
cosmology: dark matter --- galaxies: clusters --- gravitational lensing
\end{keywords}

\section{Introduction}

\subsection{The status of dark matter science}

The Cold Dark Matter paradigm (CDM) is one the of the most successful models in cosmology. The existence of a massive, non-relativistic particle that interacts only via gravity allows us to explain a wide variety of astronomical observations, including the distribution of galaxies over scales that span many orders of magnitude \citep[][e.g.]{vipers,baoBoss}. Despite the success of this model, the lack of any detection of new particles in terrestrial experiments around the weak scale means that, astronomically, 
our ability to gain further information on the particle nature of CDM further is limited \citep{LUX,xenon100}. As such we are diversifying our search, looking for new signatures that might lead us away from the CDM paradigm and give us telling insights in to its nature.

Extensions to the CDM paradigm are becoming increasingly commonplace. Models of dark matter that invoke a self-interaction \citep[][e.g.]{Harvey_BAHAMAS,Harvey_dwarf,ObserveSIDM}, or assume an ultra-light state \citep[][e.g]{FuzzDM,ultralightdm} or models that do not assume that dark matter is generated at non-relativistic velocities \citep{WDMbode,sterileNeutrinoBoyarsky,sterileNeutrinoKusenko} have become popular predicting discriminate signatures. In this letter we will study the imprint of two popular Warm Dark Matter (WDM) models on the expected flux ratios observed in strongly lensed quasars, identifying what key observation will we require in order to significantly detect or rule out these particular models.

In this paper we will concentrate on WDM. Interest in WDM has grown since the detection of an unidentified X-ray emission line in clusters of galaxies \citep{3kevline,3kevBoyarsky,3kevFranse,3kevUrban} that is consistent with a 7~keV sterile neutrino \citep{SterileNeutrino,SterileNeutrinoA,SterileNeutrinoB,SterileNeurtrinoC,SterileNeutrinoD}. Unlike CDM, WDM is produced relativistically at radiation-matter equality, generating very different growth of structure at scales with wavenumber $k>1$~$\rmn{Mpc}/h$. With higher particle velocities, WDM is able to free-stream out of the smallest scale perturbations, suppressing structure. The mass-scale of this suppression depends on the mass of the dark matter particle, $m_\chi$ \citep{sterileNeutrinoShi,sterileNeutrinoAbazajian,sterileNeutrinoAsaka,sterileNeutrinoGhiglieri,sterileNeutrinoVenumadhav}. Thus if we are able to probe the dark matter haloes down to $\sim10^8$, we will be able to constrain WDM as a plausible dark matter candidate.

Our goal in this paper is to generate an observable that will allow us to distinguish clearly between CDM and WDM. We will focus on two specific models of WDM; however, the results have much wider implications than just these models. 
Regardless of whether the dark matter is a thermal relic or is generated in some more complicated process, as long as it shares as a similar primordial transfer function to those that we use here the results will be applicable.

\subsection{Strong lensing flux anomalies} 
When the geodesics from distant sources intersect compact objects they are bent and deformed. In the rare event that the intervening compact object is dense enough it can cause the geodesics to split, resulting in multiple images of the same source. Strong gravitational lensing occurs at many different scales, for example distant galaxies can be lensed by massive foreground clusters of galaxies, which allows for the precise measurement of the total mass in their cores \cite[e.g.][]{cosmicBeast} and the study of the nature of dark matter \citep[e.g.][]{Harvey_BAHAMAS}. It is also observed on galaxy scales where background light from a distant galaxy is lensed by a foreground one, distorting the galaxy into a giant arc surrounding the foreground lens. As this light from the distant source is lensed, it is slightly perturbed by the small-scale structure along the line of sight and therefore can allow for the sensitive measurement of galaxy structure \citep[e.g.][]{galGalStrongLensingNightgale,2017MNRAS.467.3970G}, including any small substructures and hence where it may be sensitive to the dark matter model \citep{galGalLensingWDM,SL_fluxratio,compound_lens_substructure,sl_nature_dm,SLACS_SL,WDM:HE0435}. This manifests itself as small deviations in the arcs and can be directly observed.

In a similar way, distant quasars can experience the same distortion, where the light from a point source is split into discrete lines by a foreground galaxy, resulting in two or four images of the same quasar. In this situation perturbations along the line of sight cause the flux of each image to deviate from what is expected of a smooth foreground halo model. Given that the flux ratios between different images can be predicted very precisely, any anomalies in these ratios can be attributed to small-scale structure \citep{compound_lens_substructure,sl_nature_dm}.

Measurements of flux anomalies have been already been observed in quadruply imaged quasars \citep{SL_fluxratio,microlensing0924,SHARPfluxRatios}, predicting that in order to account for the data there must be $\sim5-10$~per~cent of the lensing mass in a substructure. However, recently it was shown that assuming an oversimplified initial lens model for the foreground galaxy can result in a biased estimate of the amount of substructure required to account for the data \citep{baryonFluxRatio}; similarly a baryonic disk could mimic flux anomalies \citep{diskFluxAnomalies}. As a result, in this study we adopt a method that 
takes a very different approach to the modelling of the foreground lens, which assumes only that the cosmological simulations we use are a good representation of the population of observed lensing galaxies. 
Integrating over all possible lenses, lens and source redshifts, we {would be able to} produce a prediction of what a large scale survey would observe should it observe a complete sample of multiply imaged quasars. This way observations of flux ratios can be directly compared to the simulations with no modelling required.

\section{Simulations and sample selection}\label{sec:simsamp}
In this paper we will study three models of dark matter: CDM and two models of WDM. The initial power spectrum of the WDM simulations are set by two basic particle physics parameters, namely the dark matter mass and the lepton asymmetry \citep{sterileNeutrinoShi,sterileNeutrinoVenumadhav,SterileNeurtrinoC}. Lepton asymmetry is the result of the theoretical process called leptogenesis, where lepton number, a quantity normally conserved, is instead not conserved; the lepton asymmetry quantity is defined as
\be
L_6=(n_{\nu_e}-\bar{n}_{\nu_e})/s,
\ee
where $n_{\nu_e}$ is the lepton number density, $\bar{n}_{\nu_e}$, the anti-lepton number density and $s$ the entropy density. In this study we adopt two values of $L_{6}$: $L_6=8$ and $L_6=11.2$. 
$L_6=8$ corresponds to the ``coldest'' WDM 7~keV neutrino, and $L_6=11.2$ the warmest consistent with the the resonantly produced sterile neutrino decay interpretation of the 3.55~keV line \citep{SterileNeutrinoD,sterileNeutrinoSchneider,sterileNeutrinoLovell}
, thus representing the full range parameters in the scope of this particular model. Further details on these models as applied in astronomy are available in 
\cite{sterileNeutrinoLovell}, and the matter power spectra of the runs are shown in figure~1 of \citet{lovell17b}. Hereafter we refer to these two 7~keV-mass models as L8 and L11.



The simulations in this study are the four zoomed volumes centred on the host haloes of giant elliptical galaxies introduced in \cite{lensingEAGLE}, a subset of the simulations of \citet{oppenheimer16}. A detailed description can be found there: here we present a short summary. 

The four volumes were selected to be haloes from the EAGLE project \citep{schaye15} that were: (i) suitable for resimulation at higher resolution as determined by \citet{oppenheimer16}, and (ii) suitable for lensing studies in the study of \citet{despali17b}. The CDM simulations were run for the \citet{oppenheimer16} study, and the WDM runs for the subsequent Despali~et~al.~(in prep.) paper. 

All of the simulations were run with the EAGLE galaxy formation code \citep{schaye15,crain15}, which is a heavily modified version of the {\sc gadget-3} code \citep{springel08b}. The model features pressure-entropy SPH (\citealp{hopkins13}, see also \citealp{schaller15} for further discussion), cooling, star formation, stellar evolution, supernova feedback and active galactic nuclei (AGN) feedback. The runs were performed with the RECAL version of the EAGLE galaxy formation model, which was optimized for  simulations in which the gas particle mass is approximately $2.3\times10^{5}M_{\odot}$, which is also the gas particle mass in our simulations. Haloes are identified using the friends-of-friends (FoF) algorithm and their constituent subhaloes are computed using the {\sc subfind} gravitational unbinding code \citep{springel01,dolag09}. The cosmological parameters were chosen to be consistent with the \citet{planck1_14} constraints: $h_{0}=0.6777$, $\Omega_{0}=0.307$, $\Omega_\rmn{b}=0.04825$, $\Omega_{\Lambda}=0.693$, $n_\rmn{s}=0.9611$ and  $\sigma_8=0.8288$. The WDM simulations differ from their CDM counterparts only in the application of the WDM power spectra discussed above in their initial conditions.
Finally we extract galaxies from each re-simulated halo out to $100$kpc and project in along three axes out to $1$Mpc.

The host haloes were selected by \cite{despali17b} to have halo masses, stellar masses, stellar radii and velocity dispersions that are a good match to the SLACS observational sample. We note, however, that the EAGLE simulations, plus related simulations such as ours, are subject to the equipartition problem described by \cite{ludrow2019}, which steepens the dark matter profile and expands the stellar component. Future simulations will require that this effect is taken into account.


\figs
\includegraphics[width=\textwidth]{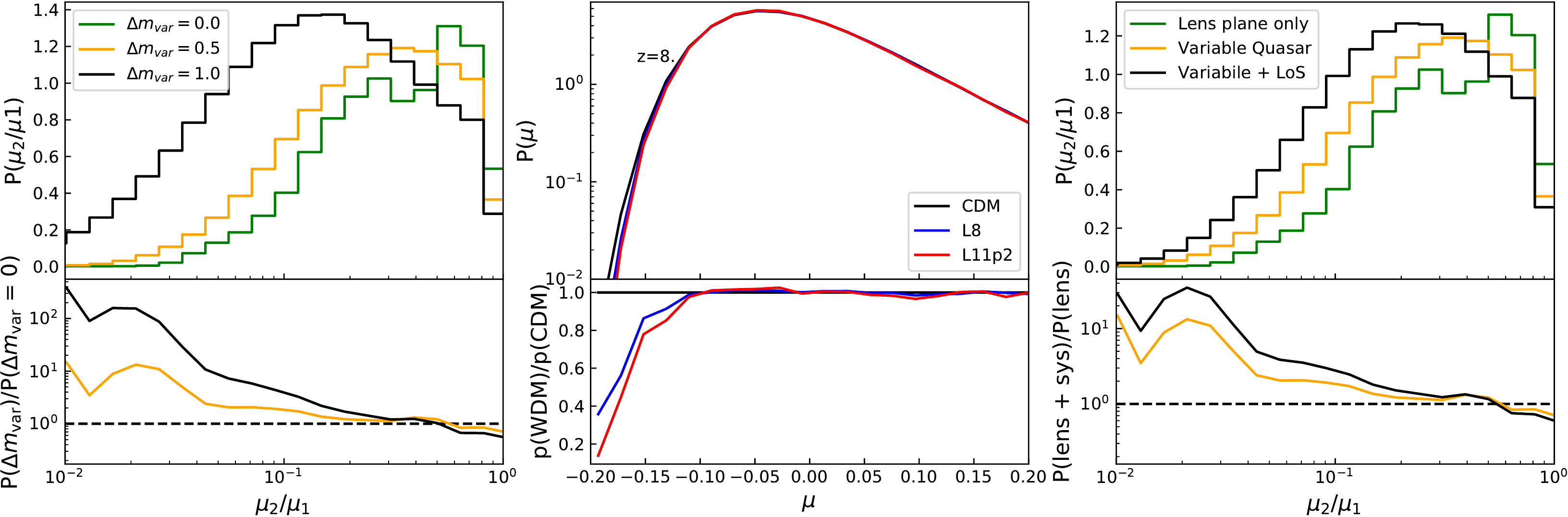}
\caption{\label{fig:turboGLpdfs} {\it Left:} The effect of quasar variability on the doubly imaged flux ratio probability density function (PDF) for the CDM case. The green line represents no quasar variability ($\Delta m=0$), the orange line  represents the expected quasar variability ($\Delta m=0.5$) and the black line strong quasar variability ($\Delta m=1.0$) as measured from COSMOGRAIL survey \citep{CosmograilDR1}  The bottom panel shows the ratio relative to $\Delta m = 0$. {\it Middle:}  The PDF of weak lensing by line of sight structures as a function of magnitude for the three cosmologies: CDM (black), L8 (blue) and L11 (red) integrated from a source redshift of $z_{\rm s}=8$. The top panel shows the absolute PDF and the bottom panel the relative difference between the WDM and CDM PDFs. {\it Right:} The PDF of the intrinsic flux ratios for a CDM halo at a redshift of $z=0.74$ (green), 
and the final convolved PDF including quasar variability (orange, $\Delta m=0.5$) and line-of-sight structures (black). The bottom panel shows the ratio of the two relative to the intrinsic distribution.}
\efigs

\section{Method}
In this letter we choose to be agnostic in our estimation of the flux ratio cumulative probability density distribution (CDF) when it comes to survey dependent attributes. However, despite this, systematics exist that are independent of survey choice that must be examined. In this section we outline how we calculate the flux ratios, specifically:
\begin{itemize}
    \item We outline the initial calculation of the flux ratio
    \item We describe how we fold in intrinsic quasar variability
    \item We describe how we incorporate perturbations from line of sight structures
\end{itemize}

\subsection{Strong lensing flux ratio calculation}\label{sec:stronglensfr}
We solve for strong lensing assuming the projected densities from simulations are at a single redshift, and hence we use the thin-lens approximation~\citep{gravitational_lensing}. We initially obtain a solution to the lens equation:
\be
\beta_i = \theta_i - \alpha_i(\theta_i), 
\ee
which is the deflection map $\alpha_i(\theta_i) $ that maps 
 lens-plane positions $\vec \theta$ to source plane positions $\vec \beta$. The solution is obtained using a Fast Fourier Transform (FFT) as a poisson solver. The deflection angle is then given by the gradient of an obtained lens potential:
\be
\alpha_i(\theta_i) = {\bf \triangledown}\Psi(\theta_i),
\ee 
where the lens potential is 
\be
\Psi =  \frac{D_{\rm LS}}{D_{\rm S}D_{\rm L}}\frac{2}{c^2}\int\Phi(D_{\rm L}, \theta_i)dz,
\ee
where the $D$ variables are the angular diameter distances between the observer, the lens (L) and the source (S), and $\Phi$ the Newtonian 3D potential.
Finally the magnification is given by the laplacian of the potential, being equal to the jacobian of the coordinate transformation:
 as a consequence of the FFT choice, we have taken care that there is sufficient empty space around the lens, in order not to get unphysical results owing to the periodic boundary conditions.
 
This deflection map then needs to be numerically inverted (using simply bi-linear interpolation), to obtain a regularly spaced sampling of the source plane positions. Specifically, we invert the vertices of the lens plane and find those cells in the source plane that lie within each image plane cell. In this way we are implicitly simulating a finite-sized source, which is by default the pixel size of the image plane ($100$pc). Although this is much larger than the size of a quasar, we are limited by the mass resolution of the simulation, and any higher resolution source size would provide no extra information in the results. As such we assume that there are no structures smaller than this $100$pc scale that would impact the flux ratios.
We therefore have at each position in the source plane, one or more lens-plane positions at which the source will be observed by the observer behind the lens.

From the inverted map, we collect all the pixels that have exactly two or three (with the central image) images on the lens plane. We discard the central images, and we discard quadruply or more lensed source pixels. The remaining data are hence our bare samples of doubly imaged source plane pixels, with their associated magnifications. In other words, we take a uniform prior on the position of a source in the source plane, by sampling all positions and assigning them the same probabilistic weight.

For a given lens at a fixed redshift $z_{\rm lens}$, we are interested in its probability distribution of flux ratios for all possible source positions. We therefore populate the source plane and then nominally integrate this plane to infinity. In practice this is impossible and therefore choose to integrate to some maximum source redshift. However, at high redshift $dV/dz$ (for volume $V$) tends to zero and hence choosing a high-redshift cut-off will not affect the results. In the presence of a actual survey, it would be trivial to integrate over the redshift distribution of the quasar sample; however, here we are agnostic about the survey choice and therefore integrate out to a redshift of $z_{\rm source}=8.0$.
Similarly, when we combine the statistics from lenses across multiple lens redshifts, we weight the distribution from each redshift according to the analytic integral over comoving volume, such that the number of flux ratios for a given lens represents the total observable volume at the redshift of that lens.

We will thus focus this paper on doubly imaged quasars. This is motivated two-fold. First, with the limited number of simulated haloes, concentrating on doubly imaged quasars means we can construct a simple observable and garner large statistics. Second, and similarly, in any large scale surveys the number of doubles will outnumber quads by a factor of $\sim20$. We found in this study that there were 12, 22 and 26 times more doubly imaged source positions than quadruply imaged ones for CDM, $L_6=8$ and $L_6=11$ respectively. The premise of this method is in measuring the flux ratios of many quasars, and therefore relies on large numbers of quasars in any future survey \citep[e.g.][]{EUCLID}. Throughout this paper we will measure the ratio between the fainter and brighter quasar image, 
\be
F =  \frac{\mu_2}{\mu_1} {\rm ~~~where~~~} \mu_1>\mu_2.
\ee

\subsection{Quasar Variability}\label{sec:quasarVariability}
It is known that quasar flux varies on timescales of $\sim 10$~days \citep{quasarVariability}. When a variable quasar is strongly lensed and the light beams are split the differential length path of each geodesic results in a time delay in each photon arrival. As a result, at any given time there may exist an anomalous flux ratio that is due to the delayed arrival of quasar variability, which may mimic small structures. In order to incorporate these into our predictions we first state that the magnitude change due to quasar variability, $\Delta m$ is
\be
\Delta m = -2.5 \log_{10}( f'/f)
\label{eqn:fluxMag}
\ee
where $f'$ is the varying flux and $f$ is the mean flux. Hence the impact on a flux ratio, $F$, would be:
\be
F' = 10^{-0.4\Delta m}\frac{\mu_2}{\mu_1} = 10^{-0.4\Delta m}F.
\ee
Following this, we can work out the impact on the probability distribution of the final flux ratio, $P_F$,
\be
P_F'(F'){\rm d}F = P_{F}(F){\rm d}F P_{\Delta m}(\Delta m) {\rm d}\Delta m,
\ee
where $P_{F}(F)$ is the flux ratio PDF in an unvarying quasar and $P_{\Delta m}$ is the PDF for magnitude change between two images. Given that:
\be
{\rm d}\Delta m = \frac{-2.5}{\log(10)F'}{\rm d}F',
\ee 
it follows that:
\be
P_F'(F') = \frac{-2.5}{\log(10)F'} \int {\rm d}F P_{F}(F)P_{\Delta M}(-2.5\log[F'/F]).
\ee
We find that the resulting probability distribution is just a convolution of the intrinsic flux ratio PDF and the PDF of the varying quasar magnitude. To calculate the final PDF we assume that the quasar variability PDF, $P_{\Delta M}$, can be modelled by Gaussian (in magnitude space), with zero mean. We calculate the final PDF for three standard deviations of $\Delta m$=0, (unvarying) $\Delta m = 0.5$ (expected) and  $\Delta m = 1$ \citep{CosmograilDR1}. Since the quasar variability will smear out any signal, a larger $\Delta m$ constitutes as a more conservative estimate of our ability to distinguish between CDM and WDM. In the left-hand panel of Figure~\ref{fig:turboGLpdfs} we show the predicted CDFs for the flux ratios of CDM for these three cases of varying quasar flux, with $\Delta m=0$ (green line, no variability), $\Delta m=0.5$ (orange line, realistic variability) and $\Delta m=1.0$ (black line, strong variability). 

\subsection{Line of sight structures}\label{sec:losturbogl}
In addition to the non-linear distortion by the lens galaxy, the light rays from a lensed quasar will also be perturbed by intervening structures that happen to lie along the line-of-sight. Unlike the distortions from the lens, these perturbations are linear and considered in the ``weak'' regime. 

Moreover, even though the lensing is strong, we are still in the regime where a point source remains a point source (\ie infinitesimally small) after lensing. As a consequence, the various sources of magnification are multiplicative, and the order of multiplication does not matter. For a single image, its observed flux $\mathcal{F}_{\rm obs}$ can be decomposed into three components,
\begin{align}
    \mathcal{F}_{\rm obs} = \mu_{\rm lens} \mu_{\rm LoS} \mathcal{F}_{\rm intrinsic},\label{eq:singleflux}
\end{align}
with $\mathcal{F}_{\rm intrinsic}$ the intrinsic flux of the source, $\mu_{\rm lens}$ the magnification due to the strong lens (computed as per Section~\ref{sec:stronglensfr}), and $\mu_{\rm LoS}$ the magnification due to the line-of-sight structure.

For the computation of the line-of-sight contribution, we will assume that there is a negligible probability of having a structure in the line of sight that is of a comparable mass to the main lens such that compound strong lensing occurs. Indeed, compound lenses are almost absent from current samples of strong lenses. In this case, all structures in the line of sight will have a smaller effect, which justifies the weak lensing approximation. Moreover we assume that since the two images of the quasar are separated by the approximately the size of a galaxy, any halo that is larger than this will affect the two fluxes equally and hence have no affect on the flux ratio. As such we therefore consider a mass function between $10^{13}M_\odot<M_{\rm los}<10^4M_\odot$. Although this mass will be redshift dependent, by taking a large mass it will smooth out any differences and cause us to overestimate the number of lenses required to discriminate. As such in this sense it is a conservative limit. 

In order to calculate the contribution of lensing by haloes along the line-of-sight we first derive a probability density distribution of magnification by large scale structure, $P_G(z_{\rm source})$. To do this we use TurboGL \citep{turboGL,turboGLa,turboGLb}, a stochastic approach to cumulative weak lensing that models the line of structure through the halo model. We modify the input mass function of TurboGL in order to simulate the effect of line of sight structures in a WDM cosmology. In order to do this we adopt the WDM correction,
\be
n_{\rm WDM}/n_{\rm CDM} = \left(1+\frac{M_{\rm s}}{M}\right)^{-\beta},
\label{eqn:wdmMassFunction}
\ee
where we fit the two free parameters $\beta$ and a characteristic mass $M_{\rm s}$ \citep{nonlinearWdm,lovellWDM} (note that this is not the equivalent half-mode mass since this shape of these two functions are very different). 
We find that 
L8 $(z=0.5)$ $M_{\rm S} = 1.6\times10^9$ and $\beta=0.35$, 
L8 $(z=0.2)$ $M_{\rm S} = 1.5\times10^9$ and $\beta=0.37$,
L11 $(z=0.5)$ $M_{\rm S} = 4.1\times10^9$ and $\beta=0.55$,
L11 $(z=0.2)$ $M_{\rm S} = 1.5\times10^9$ and $\beta=0.52$. 
We convert the CDM mass functions in TurboGL into L8 and L11 equivalents using the fits found here and calculate the weak lensing probability distribution by line-of-sight structure. The central panel of Figure~\ref{fig:turboGLpdfs} shows examples of $P_G$ at redshift $z=8$ with the ratio to CDM in the bottom panel;  the example of $z=8$ is chosen to show a maximal difference between each PDF. We note that given that the estimate of the mass function is carried out over the simulation box, which itself is an overdense region, means that we may bias our estimate of $M_{\rm s}$ and $\beta$. We therefore test the sensitivity of our final distributions to the two parameters and find that a difference of 2 orders of magnitude in $M_{\rm s}$ results in a difference of $\sim0.1$\% in the PDF.
\fig
\includegraphics[width=0.5\textwidth]{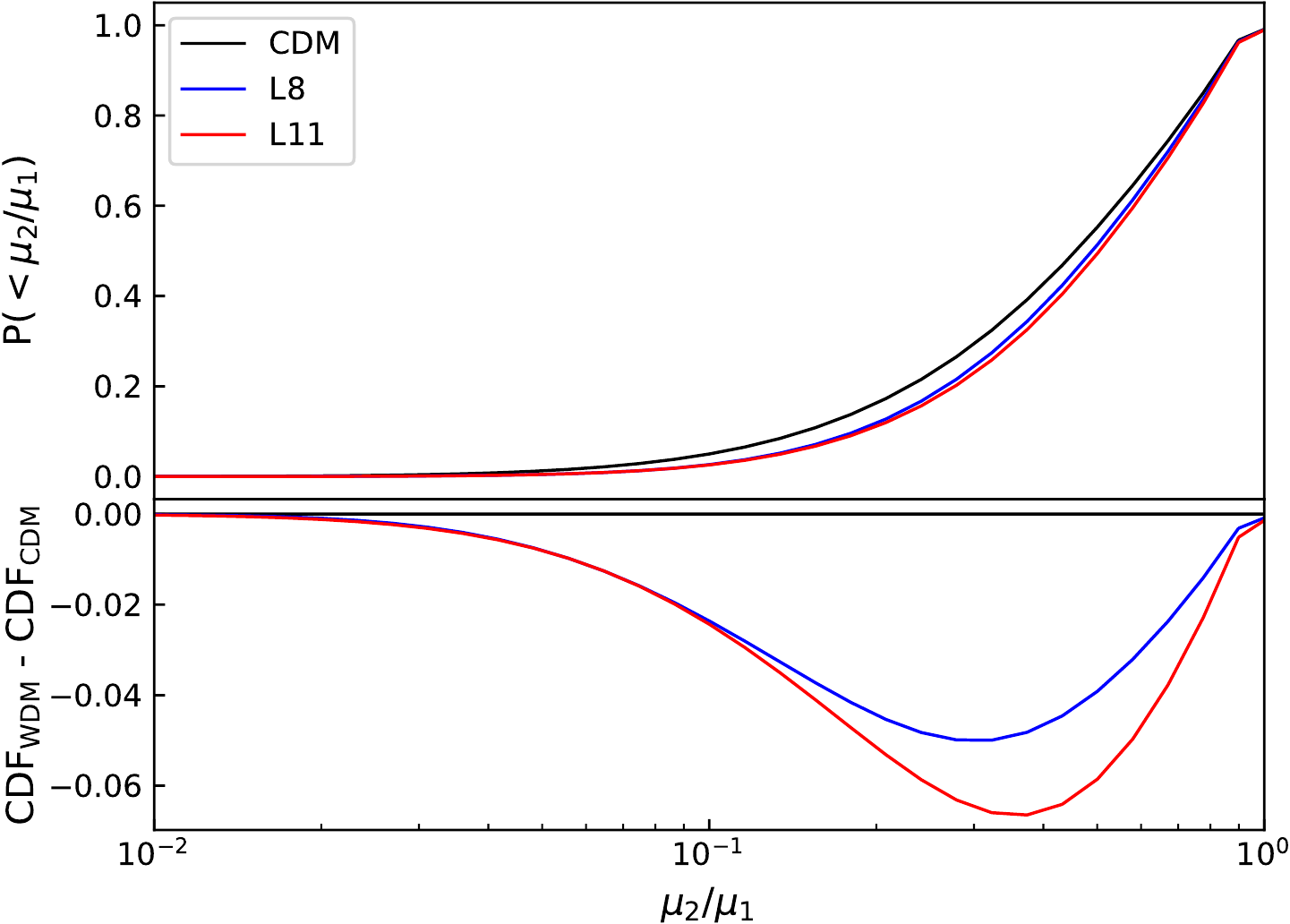}
\caption{\label{fig:AllLenses} Final cumulative distribution function of doubly imaged quasar flux ratios including the effect of line-of-sight structures and quasar variability (assuming $\Delta m=0.5$), for all lenses over five lens redshifts ($z=0.2,0.25,0.37,0.50,0.74$) for CDM (black), L8 WDM (blue) and L11 WDM (red). The bottom panel shows the difference between the two WDM models and CDM.}
\efig


\section{Total flux ratios}
Having identified the key systematics associated with the signal we now outline how we construct the final PDF.
Following equation \eqref{eq:singleflux}, we can write the flux ratio of a single pair of images as,
\begin{align}
    h\equiv \frac{\mathcal{F}_{\rm obs, 1}}{\mathcal{F}_{\rm obs, 2}} = \frac{\mu_{\rm lens, 2} \, \mu_{\rm LoS, 2}}{\mu_{\rm lens, 1} \, \mu_{\rm LoS, 1}}.
\end{align}
In the previous sections, we have so far computed two probability distributions, that of the intrinsically varying quasar flux ratio;
\be
F'\equiv \frac{\mu_{\rm lens, 2}}{\mu_{\rm lens, 1}}
\ee 
and the effect of line-of-sight structures:
\be
g\equiv \mu_{\rm LoS}.
\ee 
We need to combine these to get the final probability distribution $h$. Given that the two quasar images will have separate line-of-sight structures, the flux ratio is modified such that, 
\be 
h\equiv F' \frac{\mu_{\rm LoS, 2}}{\mu_{\rm LoS, 1}} \equiv F' \frac{g}{g'}.
\ee
Following this we construct the final PDF in h, $P_h$, in terms of the line of sight PDF, $P_g$, and the flux ratio PDF $P_f$, such that
\begin{align}
    \int {\rm d}h P_h(h) =& \int {\rm d}F' {\rm d}g {\rm d}g' P_F'(F') P_g(g) P_g(g') = 1 \\
    =& \int {\rm d}h {\rm d}g {\rm d}g' \frac{g'}{g} P_F(h\frac{g'}{g}) P_g(g) P_g(g'), \\
    P_h(h) =& \int {\rm d}g {\rm d}g' \frac{g'}{g} P_F(h\frac{g'}{g}) P_g(g) P_g(g').\label{eqn:finalConvolution_version2}
\end{align}

In words, we convolve the thin-lens flux-ratio probability distribution function (PDF) of a varying quasar (with $\Delta m = 0.5$) from Section~\ref{sec:quasarVariability} twice with the line-of-sight magnification PDF from Section~\ref{sec:losturbogl}, to obtain the final PDF for flux ratios of observed double images.

The right hand panel of Figure~\ref{fig:turboGLpdfs} shows the PDF of flux ratios for a CDM lens at a redshift $z=0.74$, where the green line represents the intrinsic flux ratio, the orange line convolved to include quasar variability (with $\Delta m=0.5$) and the black line for a variable quasar with line-of-sight perturbations. The bottom panel of this figure shows the relative difference of these PDFs, which is as high as a factor of thirty between the intrinsic lensing result and the model featuring both systematics at $\mu_{2}/\mu_{1}<0.05$.

\section{Results}
In Figure~\ref{fig:AllLenses}, we present the final cumulative probability distribution function (CDF) of the flux ratio for all doubles, over the three cosmologies CDM, L8 and L11 integrated over all lenses at five lens redshifts ($z=0.20,0.25,0.37,0.50,0.74$) that include line-of-sight structures and quasar variability of $\Delta m=0.5$. The top panel shows the absolute CDFs and the bottom panel shows the two WDM models relative to CDM. We find that WDM exhibits a $\sim 6$~per~cent difference relative to CDM, with CDM predicting many more small flux ratios than WDM. This is caused by the suppression of small scale structure that causes the larger flux ratios. 
As expected we find that the warmer model, $L11$ sees a slightly increases suppression of more extreme flux ratios.

\subsection{Suggested survey strategy}
In this study we have estimated the PDF of flux ratios in a way that is independent of 
any survey specific parameters. It is subsequently trivial to 
take the results and apply them, for example, to a given source redshift distribution of observed quasars. 
With our distributions we are able to estimate the number of flux ratios that would need to be observed in a survey of specified volume and depth in order to rule out these particularly models of WDM. 

The first step is to draw $N$ flux ratios randomly from the WDM PDF, 
draw a further $N$ flux ratios from the CDM PDF. We then calculate the likelihood of each sample, given the mean predicted by the CDM PDF, assuming a Poisson Probability Mass Function. i.e.
\be
\mathcal{L}({\rm WDM}|{\rm CDM})= \prod_{i=0}^{i=F_{\rm n}} \frac{\lambda^{k_i}_i e^{-\lambda_i}}{k_i!},
\ee
where $k_i$ is the observed number of flux ratios from a given dark matter model in the $i$th flux ratio bin,  $F_{\rm n}$ is the number of flux ratio bins and $\lambda_i$ is the expected number from CDM for the same $i$th flux ratio bin. 
Since we randomly draw from the WDM and CDM PDFs, we Monte Carlo this test 1000 times.
Following this step we compare the likelihoods by estimating the mean change in the Bayesian Information Criterion, BIC (with the same number of degrees of freedom):
\be
\Delta {\rm BIC} = -2(\ln \mathcal{L}_{\rm CDM} - \ln \mathcal{L}_{\rm WDM}).
\ee
In other words we are using Bayesian statistics to work out how large $N$ needs to be for the WDM and CDM distributions to be unambiguously different. Figure \ref{fig:numberFluxRatios} shows the predicted $\Delta$BIC for each WDM-CDM model pair. It is generally accepted that a $\Delta $BIC$>10$ is very strong preference for a model. Thus we estimate the number of flux ratios whereby the $\Delta$BIC$>10$ for three different values of quasar variability $\delta m=0$ (solid line), $\delta m=0$ (dashed line) and $\delta m=1.0$ (dotted line). We find that in the case of no quasar variability we would require $\sim 300-600$ flux ratios, in a realistic quasar variability scenario we would require almost double this, and then for a sample of strongly varying quasars we would require $\sim1000$ flux ratios. These numbers are definitely within reach of forthcoming surveys \citep[e.g.][]{EUCLID,LSST} and hence should be considered a promising test of dark matter.
\fig
\includegraphics[width=0.5\textwidth]{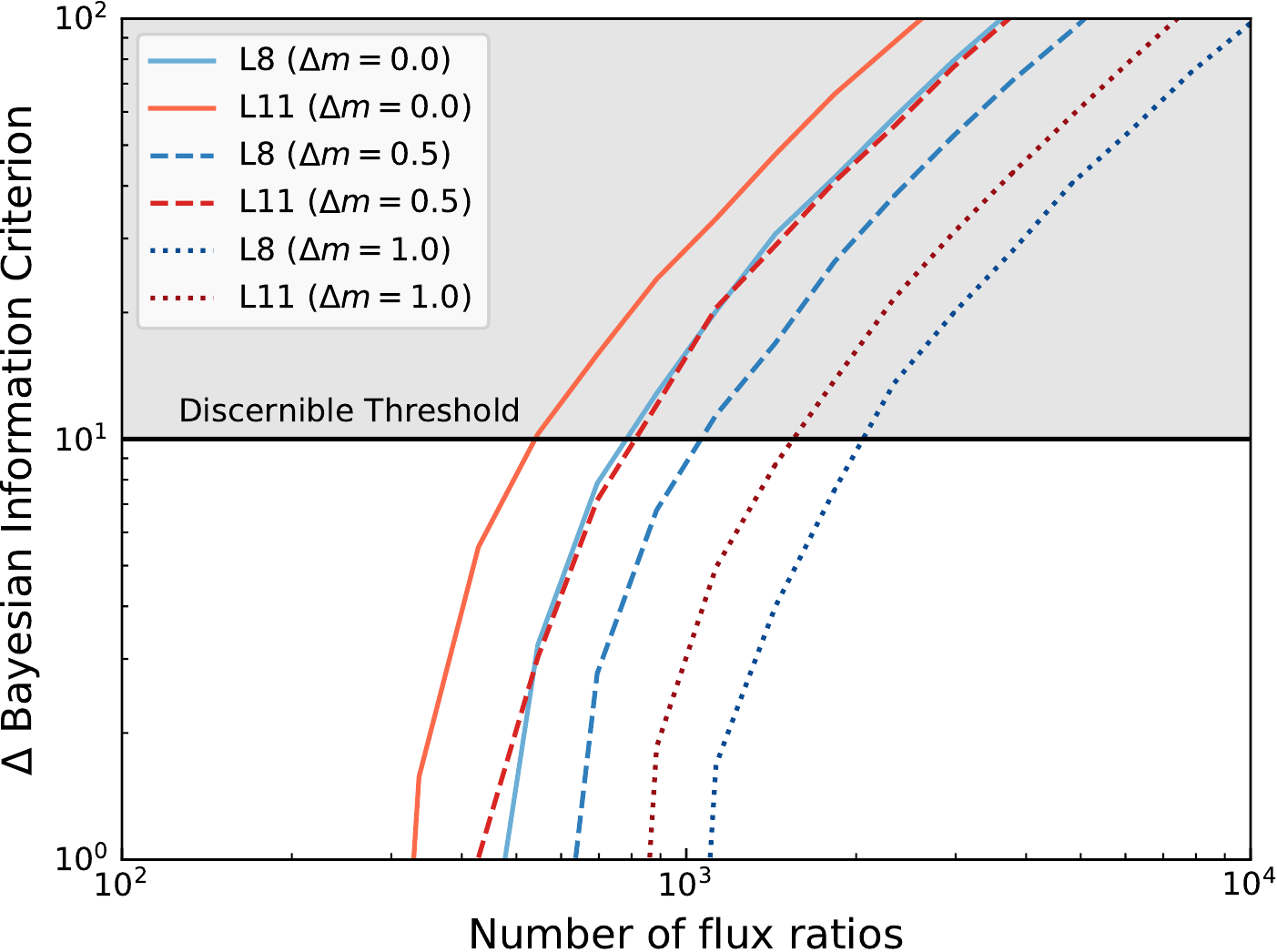}
\caption{\label{fig:numberFluxRatios} The change in Bayesian Information Criterion, an estimator for choosing between two models, for the two WDM models considered in this study. A $\Delta$BIC$>10$ is considered very strong preference for one model over another and the threshold to rule out a particular model of dark matter. Here we show the required number of flux ratios to be able unambiguously discern between either the CDM model or the paired WDM model, (L8 in blue and L11 in red) for three values of quasar variability; $\Delta m=0$ (solid line), $\Delta m=0$ (dashed line) and $\Delta m=1.0$ (dotted line).}
\efig

\subsection{Additional systematics}

In this letter we have measured the expected flux ratio PDF from double-imaged strong lensed quasars. We identified quasar variability and line of sight structures as two of the main sources of systematic uncertainty associated with this technique. From this we have estimated the number of flux ratios required to rule out the two WDM models in question here. However, there exist further systematics that still need to be addressed.

The primary theoretical systematic error will be associated with the ``modelling'' of the halos in the {\it N}-body simulations. Sub-grid modelling of the baryonic processes will affect the inner distribution of matter, altering the distribution of observed flux ratios. \cite{diskFluxAnomalies} showed that what was once interpreted as subhalos causing anomalous flux ratios, were more likely to be over-simplification of the lens models used. Here we model the {\it ensemble} distribution of lenses using state-of-the-art simulations, not individual ones. However, in the event that these lenses do not reproduce the true lenses, the comparison between observed and simulated distribution will be biased. We therefore clearly state here that in order for this statistical method to be profitable we require simulations that can produce all the known features of a true lens (e.g. baryonic disks). Hence, it will be important when comparing to observations that 
the nature of these uncertainties are understood.

Furthermore, it will be important to ensure the selection function of observed lenses matches those that are simulated. It is not clear that the selection function of observed lensed quasars is indiscriminate, and we hypothesise that a particularly type of galaxy will be more efficient at lensing than another. Therefore it may not be sufficient to integrate over all galaxies as we have done here, and thus future work will need to identify which type of galaxy, with which set of parameters, is more likely to lens a quasar and incorporate those in to these findings.

Finally, 
we have not attempted to model the contribution of dust emission and/or absorption to the lensing maps \citep{trayford15}. In the presence of dust within the lens, a single image may appear de-magnified whilst another that has no intervening dust may appear brighter, mimicking a flux ratio. Although important to be considered in observations, modelling it here is beyond the scope of this paper.

\section{Conclusions \& Discussion}
The flux ratios of strongly lensed quasars are potential probes of the nature of dark matter. Probing the small scale structure in galactic halos it has the potential to probe the power spectrum down to halo mass of $\sim10^8M_\odot$.

 Using zoom-in cosmological simulations that include realistic baryonic feedback, we have measured the distribution of observed flux ratios between the faintest and brightest image in doubly imaged quasars for cold dark matter (CDM) and two Warm Dark Matter (WDM) cosmologies.  Motivated by the unidentified X-ray line in clusters, we simulate two types of $7$~keV-mass sterile neutrino with different lepton asymmetries, $L_6=11.2$ (L11) the ``warmest'' model that can produce a 3.55~keV X-ray  emission line that is consistent with being a source of 3.5~keV line photons \citep{3kevBoyarsky,3kevline} and is consistent with the data, and $L_6=8$ (L8) the ``coldest'' possible for any 7~keV resonantly produced sterile neutrino. 
 
 We find that in our statistical approach, 
 in which the full PDF of flux ratios  
 is measured, doubly imaged quasars are $20\times$ more abundant that quadruply imaged quasars, and constitute a very efficient 
 probe of dark matter. 
 
 Once we have measured the PDF due to the intrinsic lensing signal of the lens halo, we then consider two key systematics associated with the PDF of flux ratios: quasar variability and line-of-sight structures. Folding these in, our key finding is that CDM predicts many more low image ratios than in either WDM case, with the larger difference observed in the L11 case, thus presenting a clear test for WDM. 
 
 
 Finally, we estimate the required number of observed flux ratios in order to rule out either the two dark matter models considered here or CDM. We find that in the case where the sample of quasars is moderately varying we expect a sample size of $\sim600$ observed double imaged quasars will have the statistical power to rule out both WDM models.
 
 We conclude that with the advent of large surveys, where many doubly imaged quasars will be observed with a well known selection function, it will be trivial to convolve the distributions we have found to compare the full distribution of doubly imaged quasars in different models, compare directly with observations, and subsequently to rule out WDM cosmologies, 
  having removed much of the reliance on the foreground lens model that is often cited as a source of systematic error. However, work remains to ensure that the lenses simulated are representative of the observed lens sample and that all other systematics such as dust are modelled from these simulations.

 
In conclusion, it is foreseeable that in the near future 
large scale surveys such as Euclid \citep{EUCLID} and LSST \citep{LSST} will be to measure the complete doubly imaged quasar flux ratio luminosity function and thus be able to confirm or rule out whether the unexplained 3.5~kev X-ray emission line originates from dark matter decay.

\section*{Acknowledgments}

MRL is supported by a COFUND/Durham Junior Research Fellowship under EU grant 609412 and also by a Grant of Excellence from the Icelandic Research Fund (grant number 173929−051). This work was carried out on the Dutch National e-Infrastructure with the support of SURF Cooperative. We would also like to thank Dr. Benjamin Oppenheimer for contributing his simulations of Cold Dark Matter, without which we would not have been able to complete this project. DH is supported by the D-ITP consortium, a program of the Netherlands Organization for Scientific Research (NWO) that is funded by the Dutch Ministry of Education, Culture and Science (OCW)

\bibliographystyle{mn2e}
\bibliography{bibliography,mlovell_extrabib}

\label{lastpage}

\end{document}